# Kitaev Magnetism and Fractionalized Excitations in Double Perovskite $Sm_2ZnIrO_6$

Birender Singh[1#], M. Vogl[2], S. Wurmehl[2,3], S. Aswartham[2], B. Büchner[2,3] and Pradeep Kumar[1*]

[1]*School of Basic Sciences, Indian Institute of Technology Mandi, Mandi-175005, India*

[2]*Leibniz-Institute for Solid State and Materials Research, IFW-Dresden, 01069 Dresden, Germany*

[3]*Institute of Solid State Physics, TU Dresden, 01069 Dresden, Germany*

**Abstract:** The quest for Kitaev spin liquids in particular three dimensional solids is long sought goal in condensed matter physics, as these states may give rise to exotic new types of quasi-particle excitations carrying fractional quantum numbers namely Majorana Fermionic excitations. Here we report the experimental signature of this characteristic feature of the Kitaev spin liquid via Raman measurements. $Sm_2ZnIrO_6$ is a strongly spin orbit coupled Mott insulator, where $J_{eff} = 1/2$ controls the physics, which provide striking evidence for this characteristic feature of the Kitaev spin liquid. As the temperature is lowered, we find that the spin excitations form a continuum in contrast to the conventional sharp modes expected in ordered antiferromagnets. Our observation of a broad magnetic continuum and anomalous renormalization of the phonon self-energy parameters evidence the existence of fractionalization excitations in double perovskites structure as theoretically conjectured in a Kitaev-Heisenberg geometrically frustrated double perovskite systems.

*#email id: birender.physics5390@gmail.com*
*\*email id: pkumar@iitmandi.ac.in*

Quantum magnetic materials having partially filled 5$d$ elements are subject of extensive studies as these are predicted to host exotic quantum phases. These materials open up new avenue in the field of condensed matter physics due to the presence of interesting competing interactions of similar energy scale such as on-site coulomb interaction, strong spin-orbit (SO) coupling, Hund's coupling and crystal-field splitting; involving interplay of spin, orbital, charge and lattice degrees of freedom [1-5]. Fierce competition between different degrees of freedom provides fertile ground to observe rich magnetic ground states such as Quantum Spin Liquid (QSL). The experimental realization of QSL, a state which precludes long range ordering of spins even at absolute zero temperature, is a long sought goal in physics, as they represent new states of matter. Systems with QSL as their ground state ($|GS\rangle$) are anticipated to be topologically active and to host exotic low energy quasi-particle excitations such as Majorana fermions which results from spin fractionalization, and gauge fields.

The so-called Kitaev Hamiltonian, in which the nearest neighbour Ising type interactions are bond-directional [6], is an exactly solvable model with QSL as the $|GS\rangle$. The recent search is for systems which realize these exotic states. As one example, it was proposed [7] that the strong SO coupled Mott insulators with an edge-sharing octahedral environment may give rise to a QSL as ground state. To date very few systems such as 2D honeycomb based structure $A_2IrO_3$ (A = Na, Li) and $RuCl_3$ [8-10] fulfils these criterion. However, the elusive Kitaev spin liquid state in these systems is pre-empted by the long range magnetic order at low temperature possibly resulting from the presence of interaction beyond the Kitaev model such as Heisenberg off-diagonal interactions. Interestingly, in spite of long range magnetic ordering, the existence of spin fractionalization and Majorana fermionic type excitations in these systems was inferred from the light scattering experiments via the observation of a broad continuum in Raman and inelastic neutron scattering [11-13] measurements, in contrary to the sharp magnon modes observed in a typical long range magnetically ordered systems [14-15] suggesting that the ordered $|GS\rangle$ may be proximate to a quantum phase transition into the Kitaev spin liquid $|GS\rangle$. The hunt is on to find an experimental evidence of a true Kitaev spin liquid especially in 3D systems. Recently, it has been suggested theoretically that strong SO coupled Mott insulators controlled by $J_{eff} = 1/2$ physics such as found in Iridium-based double perovskites (DP) $A_2BIrO_6$ (A = La, B = Zn, Mg) support symmetry allowed nearest neighbour interactions such as Heisenberg, Kitaev and symmetric off-diagonal exchange [16-17]. It is advocated that these 3D geometrically frustrated DP systems support Kitaev interactions as dominating exchange along with the Heisenberg interactions with A type antiferromagnetic as their $|GS\rangle$, with a total Hamiltonian given as $H = \sum_{i,j}(J_K S_i^\mu S_j^\mu + J\vec{S}_i \cdot \vec{S}_j)$ [17]. Here $J_K$ and $J$ are the Kitaev and Heisenberg exchange parameter, respectively; $\mu$ is the component of spin directed perpendicular to the bond connecting spins (i, j) and the pure Kitaev spin liquid phase (J = 0) is stable against the perturbative Heisenberg interactions. Recent neutron diffraction measurements also confirm this A type antiferromagnetic state in $La_2BIrO_6$ and $A_2CeIrO_6$ (A = Ba, Sr, B = Zn, Mg) thus supporting the existence of strong Kitaev interactions in these systems [18-21] owing to almost cancellation of the Heisenberg interactions from multiple Ir-O-O-Ir super exchange paths. It is also suggested that in these systems even long range magnetic order may be driven by highly directional exchange interactions owing to strong SO coupling and strong magneto-elastic effect.

Motivated by these exciting proposals of possible Kitaev quantum spin liquid phase in these strong SO coupled Mott insulators, we took up the inelastic light scattering (Raman) studies to probe the quasi-particle excitations in DP $Sm_2ZnIrO_6$, with $Ir^{4+}$ ($5d^5$), where the physics is governed by $J_{eff}$

= 1/2 picture. Unlike in case of 2D honeycomb structure these systems does not feature direct edge sharing IrO$_6$ octahedra though they do have varying Ir-O-O-Ir exchange paths, and these multiple exchange paths may easily support oxygen mediated electron hopping leading to significant Kitaev interactions [19-20]. The smoking gun evidence for a Kitaev kind of anisotropic coupling is hidden in the quantum spin fluctuations and may reveal itself indirectly via interacting with the photon and coupling with the underlying magnetic spectrum. Raman scattering directly probes the Majorana fermion density of states as magnetic Raman intensity of a Kitaev system is $\propto \sum_{i,j,q}(\omega-\varepsilon_{i,q}-\varepsilon_{j,q})|M_{i,j,q}|^2$, where $\varepsilon_{i,j;q}$ is Majorana spinon dispersion and $M_{i,j,q}$ is the matrix element creating Majorana excitations [22-23]. Raman scattering is a very powerful probe to capture signatures of these exotic quasi-particle excitation as it can simultaneously probe charge, spin, lattice, orbital and electronic excitations [24-27]. The double perovskite Sm$_2$ZnIrO$_6$ crystallizes in monoclinic structure ($P2_1/n$; space group No. 14), and consists of corner sharing IrO$_6$ octahedra [see Fig. 1a]; magnetic measurements suggest an AFM ordering at $T_N \sim 13$ K and tiny magnetic moments [28]. To our knowledge there is so far no report on this 3D double perovskite system probing spin excitations and exploring the possibility of anticipated spin fractionalization driven by strong SO coupling, a necessary denominator of the Kitaev spin liquid $|GS\rangle$.

Our measurements reveal a broad anomalous magnetic continuum, unlike sharp features which are hallmark of the long range magnetic ordering, that persists in the magnetically ordered state below 13 K. This anomalous evolution cannot be captured by the conventional two-magnon scattering, however, it seems consistent with the theoretical predictions for spin liquid phase. Figure 1b shows the Raman spectra at 4 K showing only the first order modes. A broad continuum with spectral weight at low energy survived till ~ 25 meV (see Fig. 1b inset, marked by the yellow color shading), and is attributed to the underlying magnetic scattering. To rule out the luminescence origin of this background, Raman measurements were carried out using two different wavelengths (see inset Fig. S1) of $\lambda = 532$nm (2.33 eV) and 633nm (1.96 eV). Detail analysis of the first as well as second order phonon modes is given in the Supplementary Information, along with the first principle density functional theory (DFT) based calculations of the phonon dispersion.

Optical phonons may provide crucial information of the underlying magnetic continuum via strong coupling between these two degrees of freedom. The asymmetric line shape of a mode near ~ 20 meV (S4, see Fig. 1b inset) is a clear indication of a Fano resonance and reflects the coupling of this mode with a broad continuum of magnetic origin. For a quantitative analysis, this mode is fitted to a Fano profile [29], $F(\omega) = I_0 (q+\varepsilon)^2 / (1+\varepsilon^2)$, where $\varepsilon = (\omega - \omega')/\Gamma$, $\omega'$ and $\Gamma$ are the phonon frequency and line-width, respectively, and $q$ define the nature of asymmetry. Figure 2 (a-c) display the temperature dependence of the mode frequency ($\omega'$) and line-width ($\Gamma$) of the phonon S4 (~20meV), schematic representation of the Eigen vectors for the S4 mode and the Fano asymmetry parameter (1/|q|). For comparison purpose, we have also plotted self-energy parameter for the strongest mode S16 at the bottom panel of Fig. 2a. Temperature dependence of the mode frequency and line-width may be captured by the anharmonic model (see supple. infor. for details) above $T_N$ ( ~ 13 K ) and below $T_N$ it shows softening attributed to the spin phonon coupling in the long-range magnetic ordered phase. Interestingly, temperature variation of $\omega'$ shows a slight jump around ~ 160-180 K and a similar jump in the frequency is also seen for other prominent modes (see supple. Infor.), signalling subtle local structural change near this temperature. The Fano asymmetry parameter (1/|q|) characterizes coupling strength of a phonon to the underlying continuum: a stronger coupling ($q \to 0$) causes the peak to be more asymmetric and in the weak coupling ($q \to \infty$) limit Fano lineshape is reduced to a Lorentzian line shape. Interestingly, the Fano asymmetry parameter ($1/|q|$) shows strong temperature dependence (see Figure 2 (c)), it has high value in the long-range ordered phase, and above $T_N$ (~13 K) continuously decreases till ~ 180 K, and thereafter remains nearly constant up to 330 K with a quite appreciable value till very high temperature (~ 180 K) suggesting the presence of active magnetic excitations far above $T_N$. We note that the Fano asymmetry parameter mapped nicely onto the dynamic spin susceptibility [discussed later, see Fig. 4b] implying that the Fano line shape is also an indicator of spin fractionalization and the increased value below 200 K may be translated to a growth of finite spin fractionalization below this temperature. We note that similar observation have been reported for the other 2D Kitaev materials RuCl$_3$, Li$_2$IrO$_3$ [12, 30].

We now focus on the temperature evolution of a broad magnetic continuum with spectral weight surviving till ~ 25 meV (see inset Fig. 1b, marked by yellow color shadings). In a pure Kitaev material, the underlying quasi-particle excitations in the form of continuum originates from Majorana fermion scattering [31-33] which survive even in the presence of Heisenberg exchange as a perturbative term [13, 19]. In Fig 3a,b, we have shown the temperature evolution of the underlying continuum. To make a quantitative estimate we extracted the intensity of this continuum as a function of temperature (see Fig. 3c). We note that the continuum loses its intensity with decreasing temperature and becomes nearly constant below ~ 150 K till ~ 20 K; and increases on entering the long range magnetic ordered phase below ~ 20 K. Our observation suggest that there is a change in the magnetic scattering below ~ 150 K hinting on development of finite entanglement between spins, also the existence of this magnetic scattering much above $T_N$ is a clear signature of frustrated magnetism. Generally, a broad magnetic continuum evolves into sharp one/two magnon peaks in long range ordered AFM systems. However, in Sm$_2$ZnIrO$_6$, there is no signature of any sharp peak suggesting that broad continuum observed here is not a conventional magnetic continuum but does have its origin in an intricate magnetic structure probably dominated by the Kitaev exchange interactions. We note that similar broad magnetic continua have been reported in other systems with QSL as their potential ground state such as herbertsmithite [34], $Cs_2CuCl_4$ [14], $RuCl_3$ and $Li_2IrO_3$ [11-12, 30]. In these systems as well, the continuum intensity gains spectral weight as spin liquid correlation start developing much above long range magnetic ordering temperature. Keeping that the magnetic continuum in quasi 2D system extends upto ~ $3J_K$, we have roughly estimated the Kitaev exchange parameter, $J_K$ ~ 10 meV from the upper cut-off energy of the magnetic continuum by taking the upper range of magnetic continuum [ ~ 25 meV ] also as ~ $3J_K$ in this system. We note that our estimated value of $J_K$ is similar to that of $RuCl_3$ ($T_N$ ~ 8-14 K) but almost half of $Li_2IrO_3$ ($T_N$ ~ 40 K).

Magnetic Raman scattering have their origin in dynamical spin fluctuations in a quantum paramagnetic phase and may provide a good measure of the thermal fractionalization of quantum spins. To further understand the evolution of this broad magnetic continuum and to identify the Majorana fermion contribution, we evaluated the dynamic spin susceptibility $\chi^{dyn}$ associated with the underlying continuum. Raman response $\chi''(\omega)$ reflects the dynamic properties of collective excitations and is obtained from the raw Raman intensity ( $Inten. \propto [1+n(\omega)]\chi''$ ) simply by dividing it by the Bose thermal factor ($[1+n(\omega)]$). To get $\chi^{dyn}$, first we find Raman conductivity $\chi''/\omega$ (see Fig. 4a for their temperature evolution) using Raman response. Then using Kramer-Kronig relation one may obtain the dynamic spin susceptibility as [35] $\chi^{dyn} = \lim_{\omega \to 0} \chi(k=0,\omega) \equiv \frac{2}{\pi} \int_0^\infty \frac{\chi''(\omega)}{\omega} d\omega$, to extract $\chi^{dyn}$, $\chi''/\omega$ was first extrapolated till 0 meV and integrated upto upper cutoff value of energy as 25 meV where $\chi''/\omega$ show no change with further increase in the energy. Fig. 4b shows the temperature evolution of $\chi^{dyn}$, we note that with decreasing temperature from 330K $\chi^{dyn}$ shows nearly temperature independent behaviour down to ~ 160-180 K, and on further lowering the temperature it progressively increases down to ~ 16 K. Interestingly, below 16 K, $\chi^{dyn}$ exhibit abrupt increase indicating the appearance of strong spin-spin correlations due to long range magnetic ordering. As in the pure paramagnetic phase, spins exhibit no correlation, as a result $\chi^{dyn}$ should be temperature independent. Most interestingly, the increased value of $\chi^{dyn}$ in the temperature range of ~ 160 K-16 K indicates the finite entanglement between spins in the paramagnetic state. Interestingly the corresponding energy scale to this temperature (~ 160 K) is comparable to the Kitaev exchange interaction ($J_K$), also around the same temperature continuum intensity as well as Fano asymmetry parameter show an anomaly suggesting this characteristic temperature corresponds to a crossover from a paramagnet to quantum paramagnet, at which spins are fractionalized due to development of short range correlation. We also tried to fit the $\chi^{dyn}$ using a power law as $\chi^{dyn} \propto T^\beta$ ($\beta = -0.45$, see solid green line in Fig 4b). Unlike a conventional antiferromagnet which show power law behaviour only below $T_N$ and saturate in the paramagnetic phase, $\chi^{dyn}$ display a power law behaviour much above the long range magnetic ordering temperature. The saturation of $\chi^{dyn}$ is a hallmark of spin gas configuration where spins do not talk to each other i.e. pure paramagnetic phase, however the power-law dependence of $\chi^{dyn}$ for double perovskite $Sm_2ZnIrO_6$ even much above the long range magnetic ordering temperature reflects the slowly decaying correlation inherent to spin liquid phase and the onset temperature (~ 160-180 K) triggered the fractionalization of spins into itinerant spinons [36-37]. This anomaly clearly suggest that the underlying magnetic continuum arises mainly from the fractionalized excitations. We note that our observation in $Sm_2ZnIrO_6$ is analogous to the observation in *quasi* Kitaev 2D, 3D honeycomb systems namely $RuCl_3$, $Li_2IrO_3$ and similar characteristics of the underlying magnetic continuum in $Sm_2ZnIrO_6$ and these systems indicate that in $Sm_2ZnIrO_6$, Kitaev magnetism may be realized.

In conclusion, our comprehensive Raman scattering study on the double perovskite $Sm_2ZnIrO_6$ evince the signature of fractionalized excitations. Anomalous evolution of the broad magnetic continuum and Fano asymmetry suggest thermal fractionalization of the spins into fermionic excitations and suggest that anticipated three dimensional geometrically frustrated non-honeycomb Iridium based double perovskite systems also realize the spin liquid state. Our results broaden the idea of fractionalized quasi-particle to a non-honeycomb based 3D geometrically frustrated Kitaev system with Heisenberg exchange as perturbation.

**References:**

[1] B. J. Kim et al., Phys. Rev. Lett. **101**, 076402 (2008).

[2] F. Wang and T. Senthil, Phys. Rev. Lett. **106**, 136402 (2011).

[3] B. J. Yang and Y. B. Kim, Phys. Rev. B **82**, 085111 (2010).

[4] C. H. Kim et al., Phys. Rev. Lett. **108**, 106401 (2012).

[5] L. Balents, Nature (London) **464**, 199 (2010).

[6] A. Kitaev, Ann. Phys. **321**, 2 (2006).

[7] G. Jackeli and G. Khaliullin, Phys. Rev. Lett. **102**, 017205 (2009).

[8] S. K. Choi, R. Coldea et al., Phys. Rev. Lett. **108**, 127204 (2012).

[9] X. Liu, T. Berlijn et al., Phys. Rev. B **83**, 220403 (2011).

[10] K. W. Plumb, J. P. Clancy et al., Phys. Rev. B **90**, 041112 (2014).

[11] L. J. Sandilands et al., Phys. Rev. Lett. **114**, 147201 (2015).

[12] A. Glamazda et al., Phys. Rev. B **95**, 174429 (2017).

[13] A. Banerjee et al., Nat. Mater. **15**, 733 (2016).

[14] R. Coldea, D. A. Tennant and Z. Tylczynski, Phys. Rev. B **68**, 134424 (2003).

[15] P. Kumar et al., Phys. Rev. B **85**, 134449 (2012).

[16] I. Kimchi and A. Vishwanath, Phys. Rev. B **89**, 014414 (2014).

[17] A. M. Cook et al., Phys. Rev. B **92**, 020417 (2015).

[18] G. Cao, A. Subedi et al., Phys. Rev. B **87**, 155136 (2013).

[19] A. A. Aczel et al., Phys. Rev. B **93**, 214426 (2016).

[20] A. A. Aczel et al., Phys. Rev. B **99**, 134417 (2019).

[21] A. Revelli et al., Phys. Rev. B **100**, 085139 (2019).

[22] B. Perreault et al., Phys. Rev. B **92**, 094439 (2015).

[23] W. Hayes and R. Loudon, Scattering of Light by Crystals (Dover Books, 2004).

[24] P. Kumar et al., Appl. Phys. Lett. **100**, 222602 (2012).

[25] B. Singh et al., J. Phys.: Condens. Matter **31**, 065603 (2019).

[26] P. Kumar et al., J. Phys.: Condens. Matter **26**, 305403 (2014).

[27] P. Lemmens, G. Guntherodt and C. Gros, Phys. Rep. **375**, 1 (2003).

[28] M. Vogl et al., arXiv:1910.13552 (2019).

[29] U. Fano, Phys. Rev. **124**, 1866 (1961).

[30] A. Glamazda et al., Nature Commu. **7**, 12286 (2016).

[31] G. Baskaran et al., Phys. Rev. Lett. **98**, 247201 (2007).

[32] J. Knolle et al., Phys. Rev. Lett. **113**, 187201 (2014).

[33] J. Nasu et al., Nat. Phys. **12**, 912 (2016).

[34] D. Wulferding et al., Phys. Rev. B **82**, 144412 (2010).

[35] Y. Gallais and I. Paul, C. R. Phys. **17**, 113 (2016).

[36] M. Hermele et al., Phys. Rev. B **77**, 224413 (2008).

[37] J. Nasu et al., Phys. Rev. Lett. **113**, 197205 (2014).

**Acknowledgment**

PK thanks the Department of Science and Technology, India, for the grant under INSPIRE Faculty scheme and Advanced Material Research Center, IIT Mandi, for the experimental facilities. Authors at Dresden thanks Deutsche Forschungsgemeinschaft (DFG) for financial support via Grant No. DFG AS 523/4-1 (S.A.) and via project B01 of SFB 1143 (project-id 247310070).

**FIGURE CAPTION:**

**FIGURE 1:** (Color online) **Crystal structure and Raman Spectra** (a) Schematic of the crystal structure of double perovskite $Sm_2ZnIrO_6$ (Pink, gray, dark yellow and red spheres represent Sm, Zn, Ir and oxygen atoms, respectively); Schematic on the right side of panel show the *ab* plane indicating Ir-O-O-Ir exchange paths with dotted red colour lines. Green and blue colour depicts Ir/$ZnO_6$ octahedra, respectively. (b) Raman spectra of $Sm_2ZnIrO_6$ at 4 K in the energy range of 7-100 meV showing only first order phonon modes; Inset shows the zoomed in view of low energy spectrum below 25 meV, yellow colour shading is guide to eye and indicates underlying continuum at 4 K, superimposed with the distinct low energy phonon modes.

**FIGURE 2:** (Color online) **Fano Resonance and Spin-phonon coupling** (a) Temperature variation of the self-energy parameters i.e. mode frequency and line-width, of the mode S4 (~ 20 meV) and the strongest mode S16 (~ 85 meV). (b) Schematic representation of the Eigen vectors for S4 mode, amplitude and direction of vibration is represented by the arrow length and sharp head, respectively. Pink, gray, dark yellow and red spheres represent Sm, Zn, Ir and oxygen atoms, respectively. Sm and oxygen atoms are exhibiting out of phase motion; modulating the bonds mediating the complex exchange interactions. (c) Temperature dependence of the Fano asymmetry parameter 1/|q| (dotted red line is guide to the eye); Inset: the asymmetry in line shape at 4 K (solid red colour line is the fitted line with Fano function and black colour is the raw spectra). Different colour background shading in panel (c) reflects the different magnetic phases.

**FIGURE 3:** (color online) **Magnetic Scattering in double perovskite Sm$_2$ZnIrO$_6$** (a) Temperature evolution of the background continuum at different temperature (4 K, 10 K and 16 K; dotted red line is guide to eye) and (b) at different temperatures between 25 to 330 K, yellow colour shading depicts the underlying continuum; (c) Temperature variation of the integrated intensity of background continuum (Solid red colour line is guide to the eye and different shading region depicts different magnetic phases).

**FIGURE 4:** (color online) **Raman Dynamic Response of Sm$_2$ZnIrO$_6$** (a) Temperature dependence of the Raman conductivity $\chi''/\omega$ in the temperature range of 4 to 330 K, and (b) Temperature dependence of the dynamic spin susceptibility $\chi^{dyn}$, extracted by using the Kramer-Kroning relation ($\chi^{dyn} \equiv \frac{2}{\pi} \int_0^{25 meV} \frac{\chi''(\omega)}{\omega} d\omega$). Solid green line is the fitted curve with power law $\chi^{dyn}(T) \propto T^{\beta}$ ($\beta = -0.45$). Background shading reflect different magnetic phase inferred from the Raman analysis.

**FIGURE 1:**

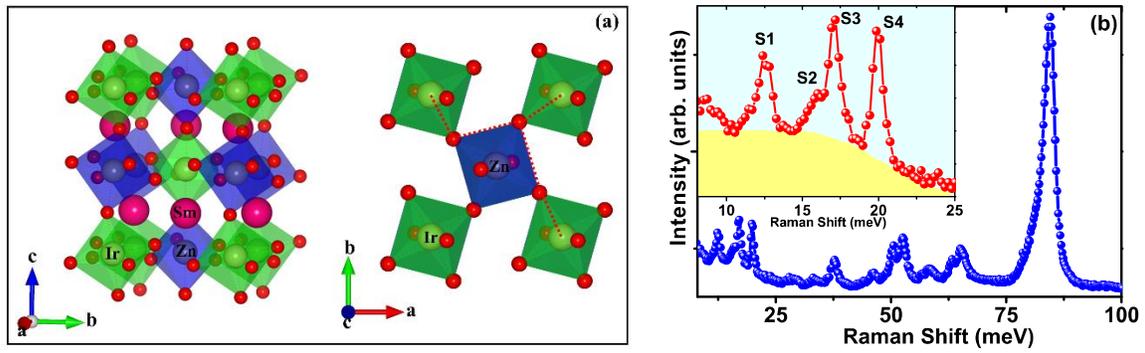

**FIGURE 2:**

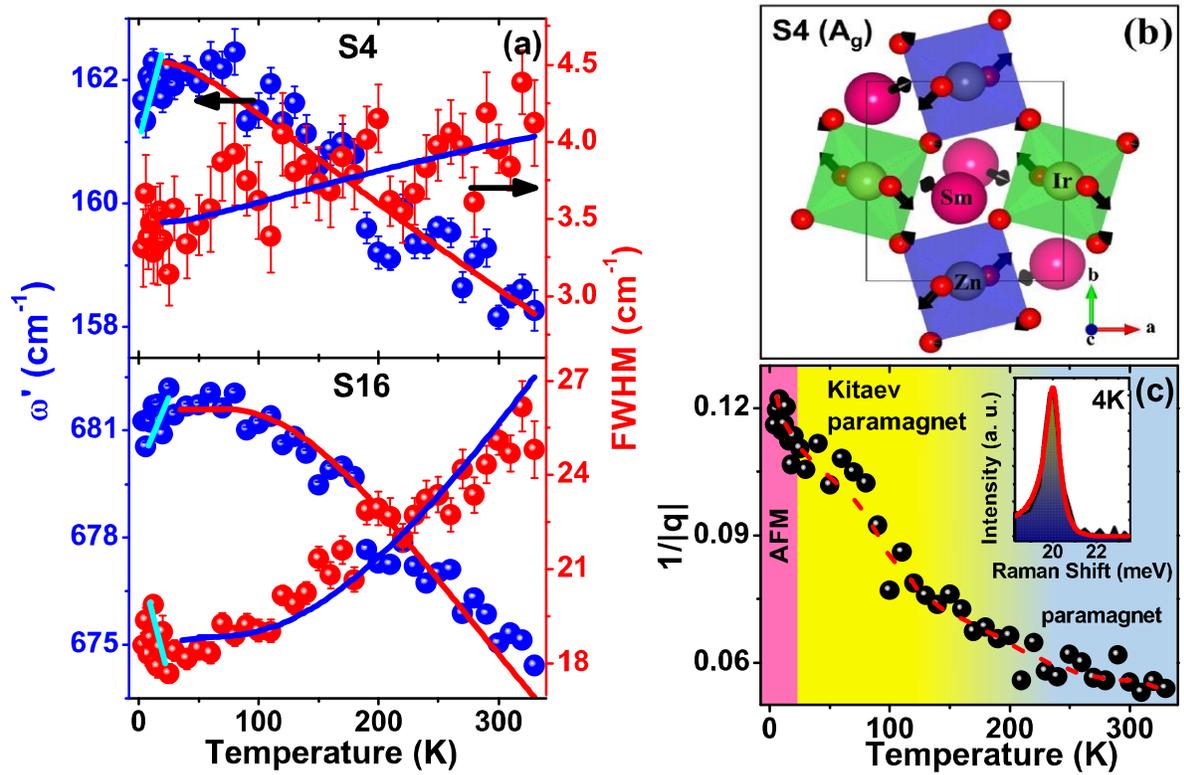

**FIGURE 3:**

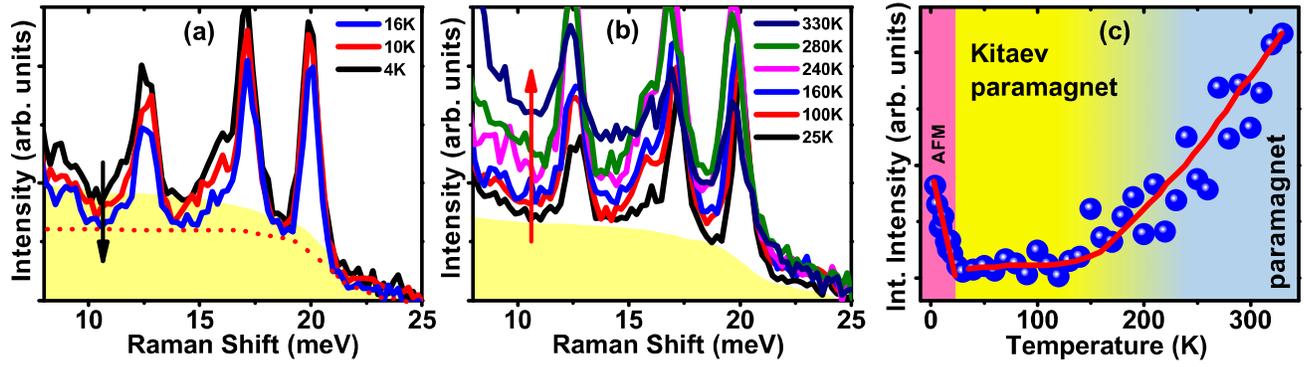

**FIGURE 4:**

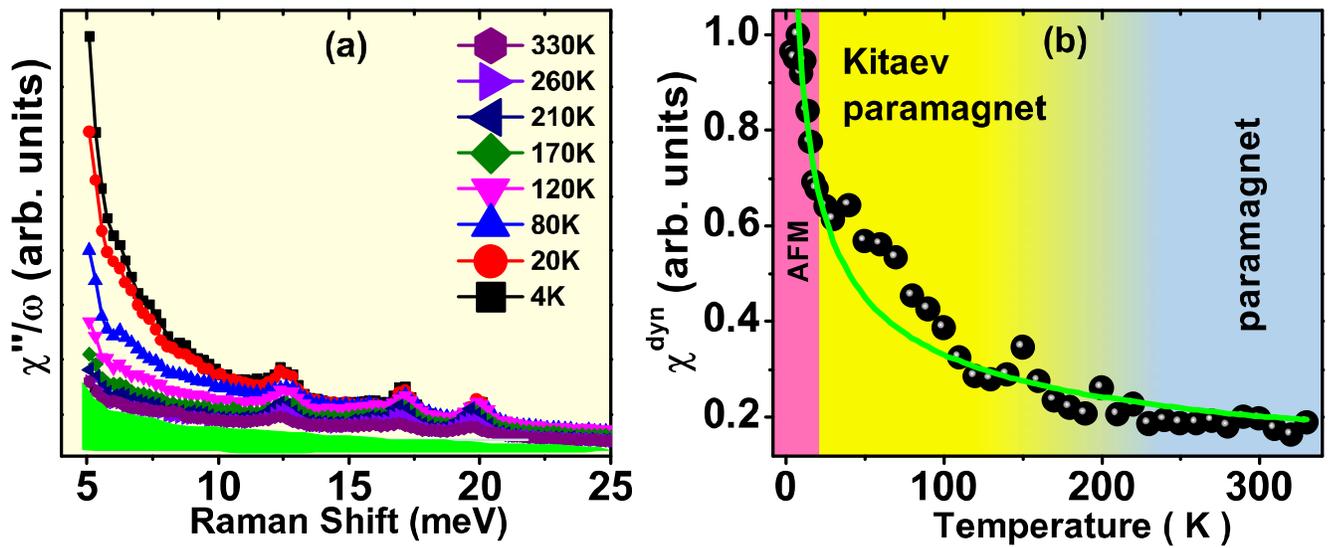

# Supplementary Information: Kitaev Magnetism and Fractionalized Excitations in Double Perovskite Sm$_2$ZnIrO$_6$

Birender Singh[1], M. Vogl[2], S. Wurmehl[2,3], S. Aswartham[2], B. Büchner[2,3] and Pradeep Kumar[1]

[1]*School of Basic Sciences, Indian Institute of Technology Mandi, Mandi-175005, India*
[2]*Leibniz-Institute for Solid State and Materials Research, (IFW)-Dresden, 01069 Dresden, Germany*
[3]*Institute of Solid State Physics, TU Dresden, 01069 Dresden, Germany*

## A. Experimental and computational details

Polycrystalline single phase Sm$_2$ZnIrO$_6$ samples were synthesized and characterized as described in Ref. [1-2]. Raman scattering experiment was done via using Raman spectrometer (Labram HR-Evolution) in backscattering geometry. Solid-state Laser $\lambda = 532 nm$ ( 2.33 eV ) was used as excitation source, focused on sample surface via 50xLWD (long working distance) objective lens (0.8 N. A.). The laser power was kept very low ($\leq 1mW$) to avoid any heating effect. The scattered light was dispersed by using 600 grooves/mm grating and a Peltier cooled charge-coupled device (CCD) detector was employed to detect dispersed light. Temperature dependent measurements in the range of 4-330 K were done via closed cycle cryostat ( Montana ), where sample was glued on top of the cold finger in an evacuated environment with ± 0.1 K or better temperature accuracy.

The first principle calculations were performed in the framework of density functional theory (DFT) via employing the plane wave method implemented in the QUANTUM ESPRESSO package [3]. We adopted generalized gradient approximation (GGA) with Perdew-Burke-Ernzerhof (PBE) as an exchange-correlation function. The dynamical matrix and phonon frequencies were calculated using density functional perturbation theory [4]. A dense $4 \times 4 \times 4$ Monkhorst-Pack grid was used for numerical integration of the Brillouin zone (BZ) [5]. The plane wave energy cut-off of 35 Ry and charge density cut off of 350 Ry was used to expand the wave function. The phonon dispersion was calculated with optimized lattice parameters as $a$ = 5.0782 Å, $b$ = 5.5070 Å, $c$ = 7.3221 Å and $\beta = 89.91^0$.

## B. Raman scattering and phonon frequencies calculations

For double perovskite $Sm_2ZnIrO_6$, group theory gives the irreducible representation of the modes as $\Gamma_{Raman} = 12A_g + 12B_g$, i.e. a total 24 Raman active modes, detail is summarized in Table-SI. Figure S1 shows unpolarized Raman response $\chi''(\omega)$ recorded in a wide energy range of 7-215 meV at 4 K showing both first as well as second order phonon modes. Raman response exhibit total 23 modes labelled as S1-S23. The spectrum is fitted with a sum of Lorentzian functions to extract a quantitative value of phonon frequency ($\omega$) and line-width ($\Gamma$). Furthermore, we performed zone centered (q=0,0,0) phonon frequency calculations to have the insight about the specific symmetry associated with a particular phonon mode and their symmetry assignment. According to our calculations phonon modes S1-S18 are attributed to be first order phonon mode, however high frequency modes S19-S23 are second ordered modes. Figure S2 (a and b) illustrate the calculated phonon dispersion curves along the high-symmetry directions and total phonon density of states (DOS) of $Sm_2ZnIrO_6$. The phonon dispersion curves shows no imaginary frequencies indicating the dynamic stability of this compound. The schematic representation of calculated Eigen vectors for zone centered phonon modes observed experimentally are shown in Figure S3. The phonon modes below 250 cm$^{-1}$ (S1-S5) are mainly attributed due to the displacement of heavy rare-earth ($Sm^{3+}$ motions) ions. Above 250 cm$^{-1}$ all the phonon modes are characterized by the motion of oxygen atom associated with Ir/ZnO$_6$ octahedra. The assignment of experimentally observed phonon modes is done in accordance to our lattice dynamics calculations listed in Table-S2.

## C. Temperature dependence of the phonon modes

Figure S4 and S5 illustrate the mode frequency and line-width of the prominent first as well as second order phonon modes as a function of temperature. The following observation can be made: (i) the phonon modes are observed to show anomalous softening ~ 1.5 % in low frequency modes ( S1-S3 ) and ~ 0.5 % in high frequency modes (S7, S9, S10 and S13) in the low temperature region with respect to its value at 16 K. (ii) Furthermore, we see the substantial line broadening (increase in phonon life-time) of phonon mode, ~ 10 % and ~ 5 % for low frequency modes ( S1and S3) and high frequency modes ( S9, S10 ) below ~ 16 K with decrease in temperature, respectively. (iii) Interestingly, we note clear and sharp discontinuity at ~ 180 K, phonon hardening ~ 2 cm$^{-1}$ in the S1, S3 and S7 modes with decrease in temperature. This discontinuity in phonon mode frequencies may occur due to local structural changes within the crystal or spin reorientations around this temperature.

To quantify the temperature dependence of phonon frequencies and damping constant, we have fitted mode frequency and line-width of first order phonon modes with anharmonic phonon-phonon interaction model in the temperature range of ~ 20-330 K. As within the harmonic approximation the phonon frequencies do not exhibit any temperature dependence and corresponding phonon life time is infinite (zero line-width). In the real system the change in phonon frequencies and finite phonon life time arises due to phonon-phonon interaction (i.e. creation and annihilation of phonons), consequently we need to take into account the anharmonic contributions in the potential term as $U_{anh}(r) = g r^3 + m r^4 + --- \equiv \beta(a^+a^+a + a^+aa) + \gamma(a^+a^+a^+a + a^+a^+aa) + ---$, where $a^+$ and $a$ are the phonon creation and annihilation operator, respectively. In the anharmonic interaction model fit we have taken both cubic (three-phonon interactions i.e. cubic anharmonicity; a decay into two acoustic phonons of equal frequency) and quadratic (four phonon interactions i.e.

quadratic anharmonicity; a decay into three acoustic phonons of equal frequency) terms given as [6],

$$\omega(T) = \omega_0 + A(1 + \frac{2}{e^x - 1}) + B(1 + \frac{3}{e^y - 1} + \frac{3}{(e^y - 1)^2})$$

and, $\Gamma(T) = \Gamma_0 + C(1 + \frac{2}{e^x - 1}) + D(1 + \frac{3}{e^y - 1} + \frac{3}{(e^y - 1)^2})$

Where $\omega_0$ and $\Gamma_0$ are the mode frequency and line-width at absolute zero temperature, respectively, and $x = \frac{\hbar \omega_0}{2 k_B T}$; $y = \frac{\hbar \omega_0}{3 k_B T}$ and A, B, C and D are constant. The solid red lines in Figure S4 are fit of the frequency and line-width to the anharmonic interaction model illustrating that in the temperature range of 20-330 K. The temperature variation of phonon modes in this range can be well explained within the cubic and quartic anharmonicities. The value obtained from fitting are summarized in Table-S2. It is worth to note that from the fitted values the cubic anharmonicity decay is primarily responsible for temperature dependences of phonon modes. The deviation of mode frequencies and line-width from the curve estimated by anharmonic interaction model is observed in low temperature regime below ~ 20 K. The phonon renormalization in the low temperature could be due to additional decay channels, namely, the interaction of phonons with other quasi-particle excitations. As at low temperature the effect of phonon-phonon interaction on self-energy parameters is minimal. Therefore, the renormalization of phonon frequency below magnetic ordering temperature can be understood due to the coupling of lattice with magnetic degrees of freedom through spin-phonon coupling. The spin-phonon coupling Hamiltonian can be written as [7]: $H = \frac{1}{2} \sum_{ij} J_{ij} (\vec{S_i} \cdot \vec{S_j})$, where $\vec{S_i} \cdot \vec{S_j}$ is spin-spin correlation function and index (i, j) runs through the lattice, $J_{ij}$ is super-exchange integral. The strength of spin-phonon coupling of particular phonon mode depends on the amplitude of vibration (atomic displacement) associated with that mode and leads to modulation of the exchange coupling. Therefore, the amount of renormalization may be different for different phonon modes. The phonon frequency is related to the spin-spin interaction by the relation given as [8-10]: $\omega \approx \omega_0^{ph} + \lambda < \vec{S_i}.\vec{S_j} >$, where $\omega$, $\omega_0^{ph}$ corresponds to phonon mode frequency and bare phonon frequency (without spin-phonon coupling), $\lambda = \frac{\partial^2 J_{ij}}{\partial u^2}$ is spin-phonon coupling coefficient ($u$ is the atomic displacement) and $< \vec{S_i}.\vec{S_j} >$ denotes the spin-spin correlation between $i^{th}$ and $j^{th}$ magnetic ions. The coupling of lattice with underlying magnetic degrees of freedom occurs due to modulation of the exchange integral $J_{ij}$ between the magnetic ions in long range ordered phase result in significant renormalisation of phonon self-energy parameters.

**REFERENCES:**

[1] M. Vogl et al., arXiv:1910.13552 (2019).

[2] M. Vogl et al., Phys. Rev. B. **97**, 035155 (2018).

[3] P. Giannozzi et al., J. Phys.: Condens. Matter **21**, 395502 (2009).

[4] J. P. Perdew et al., Phys. Rev. Lett. **100**, 136406 (2008).

[5] P. Giannozzi et al., Phys. Rev. B **43**, 7231 (1991).

[6] M. Balkanski et al., Phys. Rev. B **28**, 1928 (1983).

[7] E. Granado et al., Phys. Rev. B **60**, 11879 (1999).

[8] D. J. Lockwood and M. G. Cottam, J. Appl. Phys. **64**, 5876 (1988).

[9] E. Granado et al., Phys. Rev. Lett. **86**, 5385 (2001).

[10] P. Kumar et al., Phy. Rev. B. **85**, 134449 (2012).

**Supplementary Table S1:** Wyckoff positions and irreducible representations of phonon modes of the monoclinic ($P2_1/n$; space group No. 14) $Sm_2ZnIrO_6$ double perovskite.

| \multicolumn{3}{c}{$P2_1/n$; space group No. 14} | | |
|---|---|---|
| **Atom** | **Wyckoff site** | **Mode decomposition** |
| Sm | 4e | $3A_g + 3A_u + 3B_g + 3B_u$ |
| Zn | 2a | $3A_u + 3B_u$ |
| Ir | 2b | $3A_u + 3B_u$ |
| O(1) | 4e | $3A_g + 3A_u + 3B_g + 3B_u$ |
| O(2) | 4e | $3A_g + 3A_u + 3B_g + 3B_u$ |
| O(3) | 4e | $3A_g + 3A_u + 3B_g + 3B_u$ |

$$\Gamma_{total} = 12A_g + 18A_u + 18B_u + 12B_g$$

$$\Gamma_{Raman} = 12A_g + 12B_g \text{ and } \Gamma_{infrared} = 18A_u + 18B_u$$

**Supplementary Table S2:** List of the experimentally observed phonon mode frequencies at 4 K, fitting parameters, fitted using equations as described in the text and DFT calculated zone centered frequencies. Mode assignment (with displacement of atoms involved) has been done via our DFT based calculations. Low frequency modes, S1-S5 (< 250 cm$^{-1}$) are associated mainly with the vibrations of Sm atoms and small contribution from the vibration oxygen atoms, modes above ~ 250 cm$^{-1}$ (S6–S18) are associated with the vibration of oxygen atoms within the octahedral unit Zn/IrO$_6$.

| Mode Assignment | Exp. ω (4 K) | DFT ω | Fitted Parameters ||||||
|---|---|---|---|---|---|---|---|---|
| | | | ω$_0$ | A | B | Γ$_0$ | C | D |
| S1-B$_g$ (Sm, O) | 100.9 | 114.6 | 102.1 ± 0.3 | -0.09 ± 0.1 | -0.005 ± 0.002 | 8.7 ± 0.3 | -0.27 ± 0.14 | 0.002 ± 0.001 |
| S2-A$_g$ (Sm, O) | 126.9 | 117.2 | 128.2 ± 0.3 | -0.22 ± 0.1 | -0.003 ± 0.001 | 10.2 ± 1.0 | -0.37 ± 0.03 | -0.004 ± 0.001 |
| S3-A$_g$ (Sm, O) | 137.9 | 161.6 | 139.1 ± 0.2 | -0.32± 0.1 | -0.008 ± 0.002 | 6.7 ± 0.5 | 0.21 ± 0.1 | 0.01 ± 0.007 |
| S4-A$_g$ (Sm, O) | 160.9 | 165.6 | 162.1 ± 0.3 | -0.49 ± 0.15 | -0.01 ± 0.003 | 6.2 ± 0.2 | 0.38 ± 0.2 | 0.003 ± 0.001 |
| S5-B$_g$ (Sm, O) | 231.5 | 185.8 | 229.9 ± 1.8 | -0.12 ± 0.01 | -0.005 ± 0.002 | 19.6 ± 3.4 | 1.5 ± 0.6 | -0.05 ± 0.02 |
| S6-A$_g$ (O) | 269.6 | 302.2 | 270.5 ± 1.7 | 0.54 ± 0.3 | -0.14 ± 0.04 | 17.4 ± 2.9 | 2.8 ± 0.7 | 0.23 ± 0.07 |
| S7-B$_g$ (O) | 304.4 | 313.7 | 307.9 ± 0.6 | -2.4 ± 0.5 | -0.03 ± 0.01 | 10.1 ± 1.1 | 3.5 ± 0.9 | -0.06 ± 0.003 |
| S8-A$_g$ (O) | 370.5 | 362.1 | 376.9 ± 1.5 | -5.1 ± 1.2 | 0.1 ± 0.04 | 17.1 ± 3.7 | 2.29 ± 0.7 | -0.19 ± 0.02 |
| S9-B$_g$ (O) | 406.6 | 409.2 | 410.2 ± 1.1 | -3.26 ± 1.2 | -0.05 ± 0.01 | 10.5 ± 2.9 | 0.69 ± 0.2 | 0.9 ± 0.20 |
| S10-B$_g$ (O) | 424.6 | 431.8 | 429.2 ± 0.9 | -4.3 ± 1.0 | 0.07 ± 0.02 | 10.1 ± 2.5 | 4.95 ± 2.6 | -0.04 ± 0.01 |
| S11-A$_g$ (O) | 470.5 | 453.6 | 478.1 ±02.3 | -5.9 ± 2.4 | -0.02 ± 0.001 | 29.6 ± 5.3 | 1.7 ± 0.8 | 0.6 ± 0.10 |
| S12-A$_g$ (O) | 509.0 | 519.3 | 511.9 ± 1.9 | -2.08 ± 1.6 | 0.10 ± 0.03 | 5.3 ± 1.6 | 2.5 ± 0.7 | 1.7 ± 0.80 |
| S13-A$_g$ (O) | 526.9 | 525.8 | 533.7 ± 2.7 | -6.1 ± 2.6 | -0.06 ± 0.02 | 16.7 ± 4.9 | 3.6 ± 1.6 | -0.5 ± 0.20 |
| S14-B$_g$ (O) | 564.9 | 559.7 | - | - | - | - | - | - |
| S15-B$_g$ (O) | 662.9 | 661.2 | 667.8 ± 5.2 | -5.9 ± 1.3 | -0.9 ± 0.04 | 22.2 ± 3.8 | 16.8 ± 4.7 | -0.9 ± 0.30 |
| S16-A$_g$ (O) | 681.3 | 692.6 | 692.7 ± 4.6 | -10.9 ± 0.2 | -0.4 ± 0.12 | 5.3 ± 2.1 | 13.8 ± 3.9 | -0.08 ± 0.03 |
| S17-B$_g$ (O) | 785.6 | 784.5 | - | - | - | - | - | - |
| S18-A$_g$ (O) | 859.4 | 818.7 | - | - | - | - | - | - |
| S19 | 1131.9 | - | - | - | - | - | - | - |
| S20 | 1215.1 | - | - | - | - | - | - | - |
| S21 | 1321.2 | - | - | - | - | - | - | - |
| S22 | 1465.7 | - | - | - | - | - | - | - |
| S23 | 1558.7 | - | - | - | - | - | - | - |
| * 167.7 (B$_g$), 332.1 (B$_g$), 349.6 (A$_g$), 529.5 (B$_g$), 657.4 (A$_g$) and 703.4 (B$_g$) |||||||||

* List of calculated frequencies which are not observed experimentally. Units are in cm$^{-1}$.

**FIGURE CAPTION:**

**FIGURE S1:** (Color line) Raman response $\chi''(\omega)$ of double perovskite $Sm_2ZnIrO_6$ recorded at 4 K in a wide energy range of 7-215 meV. Inset shows raw Raman data recorded at room temperature with 532nm (2.33eV) and 633nm (1.96eV) excitation laser.

**FIGURE S2:** (Color online) Calculated phonon dispersion curves along high-symmetry directions (a), and (b) Total phonon density of states of $Sm_2ZnIrO_6$ (Inset: First Brillion zone with high-symmetry directions in red colour).

**FIGURE S3:** (Color online) Calculated Eigen vectors for all the experimentally detected (first-order) phonon modes. Pink, Gray, dark yellow and red spheres represent Sm, Zn, Ir and oxygen atoms, respectively. Black arrows on the atoms represent the direction of characteristic displacement and the size of the arrow represent the magnitude of vibration. *a*, *b* and *c* represent the crystallographic axis.

**FIGURE S4:** (color online) Temperature variation of mode frequency and line-width of S1, S3, S7, S9, S10 and S13 modes. Solid red lines are fitted curve as described in the text and yellow lines are guide to eye. Inset of S1, S3 and S7 shows zoomed view of frequency in temperature range of 60 K to 300 K and solid red color vertical line is guide to eye.

**FIGURE S5:** (color online) Temperature variation of mode frequency and line-width of the second order phonon modes S20, S21, S22 and S23.

**FIGURE S1:**

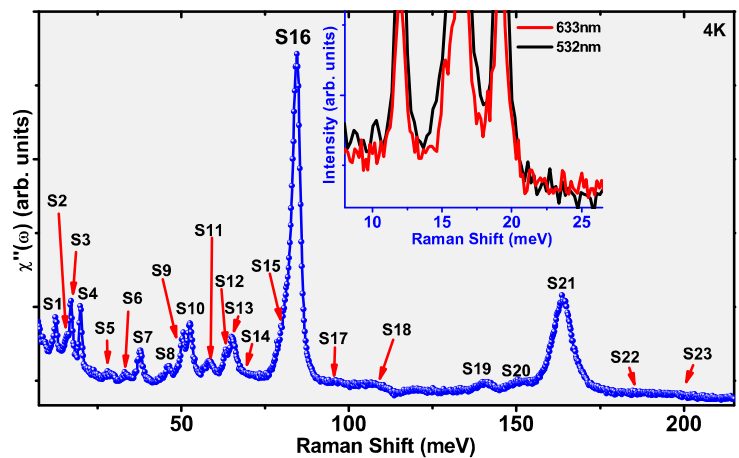

**FIGURE S2:**

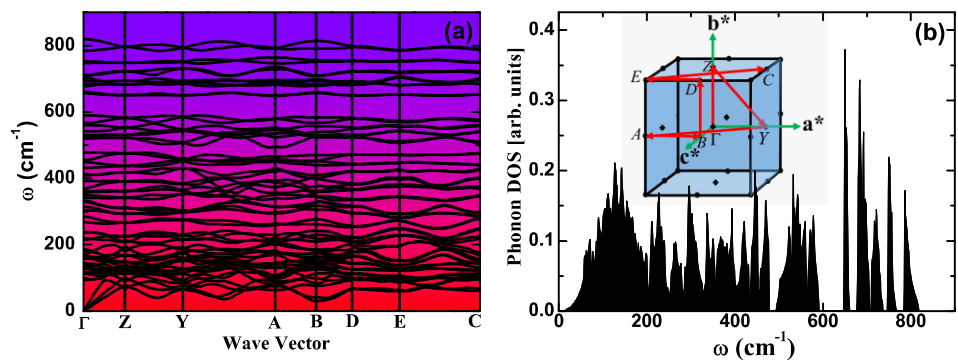

**FIGURE S3:**

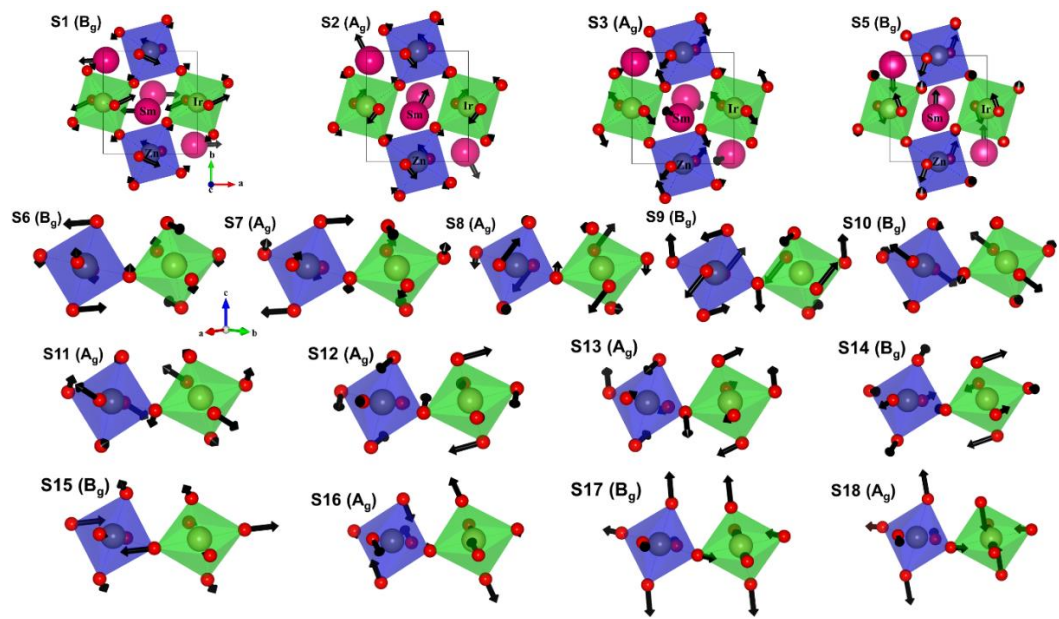

**FIGURE S4:**

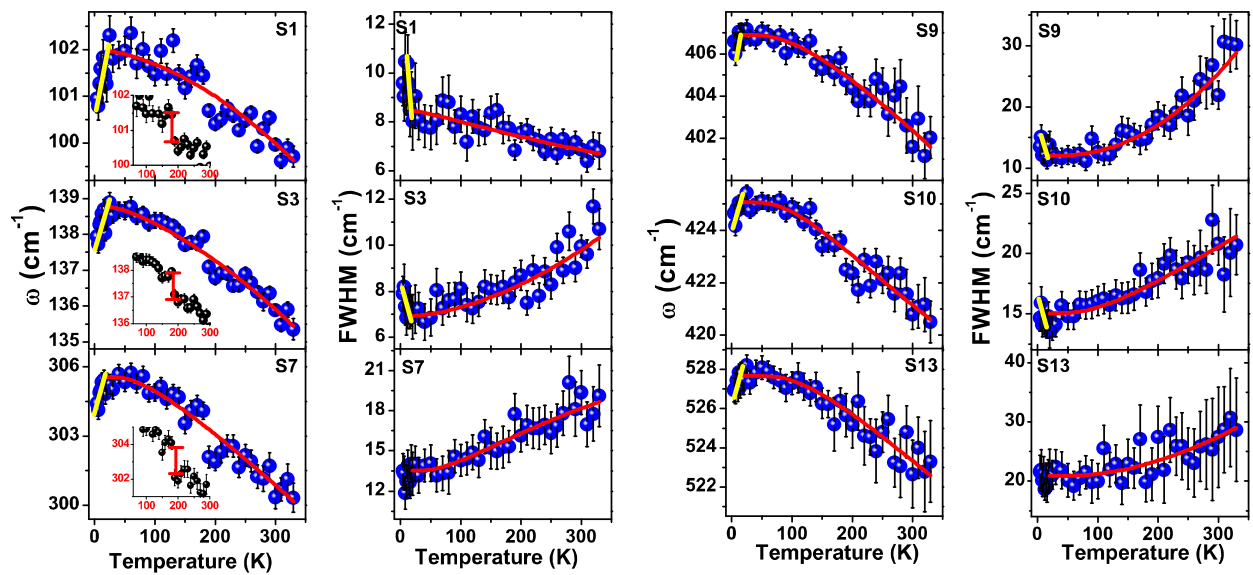

**FIGURE S5:**

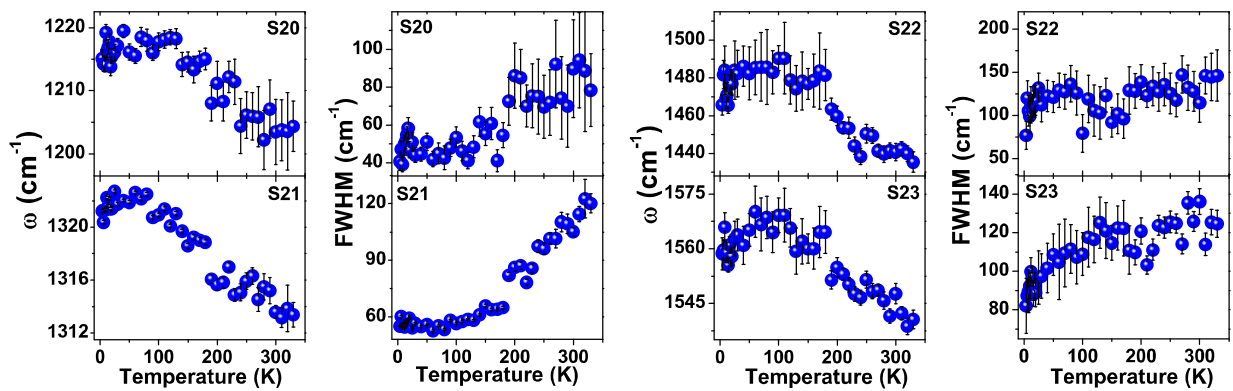
26